\documentclass[12pt]{iopart}
\usepackage{graphicx}
\usepackage{epstopdf} % Converts EPS figures on the fly
\usepackage{dcolumn}% Align table columns on decimal point
\usepackage{bm}% bold math
\usepackage{amssymb}
\usepackage{color}
\begin{document}
\bibliographystyle{unsrt}

\title{Causality detection and turbulence in fusion plasmas}

\author{B.Ph.~van Milligen$^1$, G.~Birkenmeier$^2$, M.~Ramisch$^3$, T.~Estrada$^1$, C.~Hidalgo$^1$, A.~Alonso$^1$} %D.E.~Newman$^4$
 \address{$^1$Asociaci{\'o}n EURATOM-CIEMAT para Fusi{\'o}n, Avda.~Complutense 40, 28040 Madrid, Spain}
 \address{$^2$Max Planck Institute for Plasma Physics, EURATOM Association, Boltzmannstr. 2, 85748 Garching, Germany}
 \address{$^3$Institut f\"ur Grenzfl\"achenverfahrenstechnik und Plasmatechnologie, University of Stuttgart, 70569 Stuttgart, Germany}
% \address{$^4$Department of Physics, University of Alaska, Fairbanks, Alaska 99775-5920, USA}
% \date{\today}

%===========================
\begin{abstract}
This work explores the potential of an information-theoretical causality detection method for unraveling the relation between fluctuating variables in complex nonlinear systems.
The method is tested on some simple though nonlinear models, and guidelines for the choice of analysis parameters are established.
Then, measurements from magnetically confined fusion plasmas are analyzed.
The selected data bear relevance to the all-important spontaneous confinement transitions often observed in fusion plasmas, fundamental for the design of an economically attractive fusion reactor.
It is shown how the present method is capable of clarifying the interaction between fluctuating quantities such as the turbulence amplitude, turbulent flux, and Zonal Flow amplitude, and uncovers several interactions that were missed by traditional methods.  
\end{abstract}
\pacs{05.45.Tp,07.05.Kf,52.35.Ra,52.55.Hc}
%Version: \today ; \thistime

\maketitle

%===========================
\section{Introduction}\label{introduction}

The problem of determining the causal relationship between various interacting fields or variables is of fundamental importance in many branches of science. 
Knowledge of the causal connection between variables is helpful for the elaboration of a realistic physical model and/or to check its validity. 
If one can intervene in the system under study and modulate the value of one variable, the observation of the (delayed) reaction of other variables to this modulation sometimes allows establishing the causal relationship. 
However, for various reasons many systems do not permit such intervention, or they are too complex to allow a straightforward interpretation of the observations. 
{Other techniques to uncover causal relations are based on finding precursor events, time delays between extreme events (conditional averaging), correlations, etc. One may also attempt to match the system evolution to the predicted evolution from an analytic or numerical model, or to quantify parameters related to system evolution (growth rates, damping rates, etc.).
Most of these methods, however, do not provide a direct quantification of the causal interaction between variables. 
Even worse, linear analysis techniques (correlations, conditional averages) may lead to confusing or even erroneous conclusions regarding causality (cf.~the well-known adage {\it `correlation does not imply causation'}).}
In this situation, how must one then determine the causal relation between variables?

Causality is notoriously hard to define in general~\cite{Hlavackova:2007}.
In the present work, we do not use the term `causality' in its philosophical, absolute sense (if $Y$ occurs, then $X$ will occur; or: if $X$ occurs, then $Y$ must have occurred). 
Rather, we turn to the concept of `quantifiable causality' introduced by Wiener~\cite{Wiener:1956} (rephrased slightly): 
{\it For two simultaneously measured signals $X$ and $Y$, if we can predict $X$ better by using the past information from $Y$ than without it, then we call $Y$ causal to $X$.}
This idea led to the formulation of an algorithm for the detection of the causal relation between two measured signals, denoted by {\it Granger causality}~\cite{Granger:1980}. This algorithm, however, is based on a linear prediction of the evolution of a time series involving multivariate minimization, which is inadequate for the analysis of turbulence.
Although non-linear generalizations are possible and have indeed been elaborated~\cite{Hlavackova:2007}, here we turn to a non-parametric procedure for causality detection originating in the field of information theory: the `Transfer Entropy'~\cite{Schreiber:2000}.

In this work, we are mainly concerned with the interaction between Zonal Flows and turbulence.
This interaction underlies the spontaneous confinement transitions often observed in fusion plasmas, fundamental for the design of an economically attractive fusion reactor.
In recent years, more or less detailed models for this interaction have become available~\cite{Diamond:2005}.
Much effort has been invested in demonstrating the relevance of these models for describing the observations, applying advanced analysis tools such as the bicoherence~\cite{Moyer:2001,Tynan:2001,Nagashima:2006,Fujisawa:2009}

An important aspect of these studies is the elucidation of the causal relation between turbulence, fluctuating Zonal Flows, and steady state Sheared Flows.
This issue is often implicitly present in the relevant publications, although causality is usually treated with some respect due to the difficulty of addressing it directly; usually, all that can be said is that a certain sequence of events is observed ($Y$ happens before $X$), which is a necessary but insufficient condition for the existence of a causal relation.
Some `traditional' methods for elucidating the causal relation between Zonal Flows and turbulence in fusion plasmas are~\cite{Burrell:1999}: 
(a) looking for spatial and temporal correlations (i.e., $Y$ happens before $X$)  between $E \times B$ shear, turbulence levels, and transport;
(b) comparing growth and damping rates ($\gamma_{\max}$ vs.~$\omega_{E \times B}$)~\cite{Tynan:2013} or energy transfer rates~\cite{Manz:2012b};
(c) controlling $E_r$ and the $E \times B$ shear externally and observing the effect~\cite{Xu:2009}, and 
(d) looking for `precursors' of a confinement transition~\cite{Askinazi:2012}.
These methods often rely on observing variations on a slow time scale by averaging out the fast time scale of turbulent fluctuations.
However, the confinement transition relies, precisely, on an interaction between the slow {\it and} fast time scales, and it may well be that this averaging operation eliminates essential information, thus precluding the clarification of causal relationships.

One analysis method that avoids this issue (of averaging out information) is the calculation of the energy transfer involved in quadric three-wave coupling, based on the bispectrum~\cite{Ritz:1989,Xu:2012}.
This energy transfer reveals the direction of energy flow in Fourier space.
However, bispectral techniques rely on a number of rather strong assumptions such as weak turbulence with dominant quadratic interactions and a stationary state, and requires rather lengthy time series for reliable results.
By contrast, the approach discussed here, based on the Transfer Entropy, is more generic and less dependent on underlying assumptions.

The goal of this work, therefore, is to study whether the Transfer Entropy technique can provide an answer to the causality questions of the type `which variable influences which other' in a highly non-linear situation characterized by various coupled variables or fields in a magnetically confined plasma.
For this purpose, we will analyze a few model systems considered relevant and analyze experimental data from the TJ-K and TJ-II stellarators.
The data are selected for their relevance to the study of the interaction between Zonal Flows~\cite{Diamond:2005} and turbulence. 

The structure of this paper is as follows.
In Section \ref{method}, we outline the method and perform some tests on relevant models of systems with non-linearly interacting variables.
In Section \ref{results}, we present some analysis results for data obtained in fusion devices, relevant in the framework of Zonal Flows and confinement transitions.
Finally, in Section \ref{discussion}, we discuss the findings and draw some conclusions.

%===========================
\section{Method}\label{method}

Consider two processes $X$ and $Y$ yielding discretely sampled time series data $x_i$ and $y_j$.
The data are assumed to correspond to a stationary state; 
any slow drifts of measurement signals have been removed by subjecting the time series to a suitable trend removal, if necessary.
Their Mutual Information is defined as \cite{Abarbanel:1993}:
\begin{equation}\label{mutualinfo}
I(X;Y) = \sum_{i,j}{p(x_i,y_j) \log_2 \frac{p(x_i,y_j)}{p(x_i)p(y_j)}}
\end{equation}
where $p$ is a (joint) probability distribution function (pdf).
It quantifies the mutual reduction of uncertainty of one of the variables due to knowledge of the other one, expressed in amount of bits.
The Mutual Information $I(X;Y)=0$ if and only if $X$ and $Y$ are statistically independent, in which case $p(x_i,y_j)=p(x_i)p(y_j)$.
Thus, the Mutual Information detects common information content between the processes $X$ and $Y$ but does not reveal the direction of information flow (if any). For this purpose, the temporal structure of the data patterns must be taken into consideration.

We introduce the multi-indices $\alpha = (\alpha_1, \alpha_2, \dots, \alpha_k)$ and $\beta = (\beta_1, \beta_2, \dots, \beta_l)$, such that the indices $\{\alpha_i,\beta_j\} \in \mathbb{N}$ are monotonically increasing, i.e., $0 \le \alpha_1< \alpha_2< \dots< \alpha_k$ and similar for $\beta$.
We will use the shorthand notation $x_n^{(k)} = (x_{n-\alpha_k}, \dots, x_{n-\alpha_1})$  to indicate a set of $k$ data values preceding or coinciding with the time associated with time index $n$, and likewise $y_n^{(l)} = (y_{n-\beta_l}, \dots, y_{n-\beta_1})$.
A measure of information transfer between the two time series $X$ and $Y$ is given by the {\it Transfer Entropy}~\cite{Schreiber:2000}:
\begin{equation}\label{TE}
T_{Y \to X} = \sum{p(x_{n+1},x_n^{(k)},y_n^{(l)}) \log_2 \frac{p(x_{n+1}|x_n^{(k)},y_n^{(l)})}{p(x_{n+1}|x_n^{(k)})}}
\end{equation}
{The sum runs over the arguments of the probability distributions (or the corresponding bins, cf.~next section).}
The reason for using multi-indices (a minor extension of~\cite{Schreiber:2000}) is to allow the possibility of including various time scales of influence on the effect variable.
The Transfer Entropy can be rewritten in the form of a Conditional Mutual Information~\cite{Hlavackova:2007}. It measures the excess amount of bits needed to encode the information of the process $X$ at time point $n+1$ with respect to the assumption that this information is independent from $Y$. 
In other words, the Transfer Entropy is an implementation of Wiener's `quantifiable causality'.
If $Y$ has no influence on the immediate future evolution of system $X$, one has 
${p(x_{n+1}|x_n^{(k)},y_n^{(l)})}={p(x_{n+1}|x_n^{(k)})}$, so that $T_{Y \to X} =0$.
$T_{Y \to X}$ can be compared to $T_{X \to Y}$ to uncover a net information flow.

Using $p(x|y) = p(x,y)/p(y)$, Eq.~(\ref{TE}) can be rewritten as
\begin{equation}\label{TE1}
T_{Y \to X} = \sum{p(x_{n+1},x_n^{(k)},y_n^{(l)}) \log_2 \frac{p(x_{n+1},x_n^{(k)},y_n^{(l)})p(x_n^{(k)})}{p(x_n^{(k)},y_n^{(l)})p(x_{n+1},x_n^{(k)})}}
\end{equation}
Thus, computing $T_{Y \to X}$ requires estimating four multi-dimensional probability distributions.

%{
%\subsection{Error estimate [section to be removed]}
%{\it NOTE: Schreiber does not consider defining an error measure for $T$. Also, I do not know if this is easy to do -- but it would be very valuable... here go some thoughts:}
%
%The probability distributions appearing in Eq.~(\ref{TE1}) are estimated by subdividing the ranges of the $x$ and $y$ variables in $m$ discrete bins. For simplicity, we will use the same number of bins for all variables. 
%A given bin (in the multi-dimensional space of the corresponding probability distribution) will receive a certain number of `hits', say $n$.
%The corresponding value $p$ for this bin is obtained as $p=n/N$, where $N$ is the length of the data arrays ($x,y$).
%As $n\ge0$ and $n$ is discrete, the statistical distribution of $n$ should approximately follow a Poisson distribution.
%Then, the error of a given value $n$ can be estimated as $\sqrt{n}$. 
%[But error propagation is not going to work as the $p$'s in Eq.~(\ref{TE1}) are not independent... Maybe use a Bayesian approach? Monte Carlo?]
%}

\subsection{Implementation}

Here, the probability distributions appearing in Eqs.~(\ref{TE}),(\ref{TE1}) are calculated using a discrete binning of $m$ bins in each coordinate direction. The main joint pdf $p(x_{n+1},x_n^{(k)},y_n^{(l)})$ has $l+k+1$ dimensions, so there are $m^{l+k+1}$ bins, and this number should be much smaller than the available length of the data arrays, $N$, in order to obtain a statistically significant sampling of the pdf. 

In plasma physics applications, the available stationary time series are usually rather short ($N \simeq 10^3-10^5$). 
This inevitably means that $m$, $l$, and $k$ should all be small. Choosing a small $m$ value (2 or 3) is called `coarse graining'.
For the same reason, we will set $l=k=1$ in this work. 
This means that $\alpha$ and $\beta$ are scalar indices instead of vector indices, and that their value must be chosen judiciously in order to capture the historic information that has the most significant impact on the future evolution of the signals.

This introduces the problem of selecting an appropriate value of $\alpha,\beta$. If the signals are oscillatory, the period of oscillation can be determined and $\alpha$ and $\beta$ should correspond to a time interval less than about one quarter of the oscillation period. When the signal is not clearly oscillatory or has multiple oscillations, the (linear) decorrelation time can be used as a guide for choosing $\alpha$ and $\beta$. In strongly chaotic, nonlinear, or turbulent systems, it is probably better to use the Mutual Information to determine this value, as described in Ref.~\cite{Abarbanel:1993}.
To do so, one replaces $Y$ in Eq.~(\ref{mutualinfo}) with a delayed version of $X$ and computes $I(X_t;X_{t+\tau})$ for a set of delay times $\tau$; 
in the following, we shall refer to this quantity as the self-Mutual Information.
In any case, the Transfer Entropy results should not be excessively sensitive to the precise choice of $\alpha$ and $\beta$ provided the preceding guidelines are followed, which can be tested by varying their values and observing the outcome.

%{[To be removed:]
%I have implemented the following Matlab subroutine:
%\begin{quote}
%\begin{verbatim}
%function T = transfer_entropy(x,y,lags,orderx,ordery,nbin)
%\end{verbatim}
%\end{quote}
%Here, \texttt{lags} is an array of lag values (although currently, \texttt{lags} can only be a scalar, and by default \texttt{lags} $=1$). 
%Likewise, \texttt{orderx} $ = \alpha$ and \texttt{ordery} $ = \beta$ are arrays.
%\texttt{nbin} $=m$ is the number of bins (in each coordinate direction) for the construction of the multidimensional pdf.
%[Equivalent Fortran routine is also available.]
%}

In the remainder of this chapter, we will perform some tests using well-understood though non-trivial models.

\clearpage
%===========================
\subsection{A system of coupled Van der Pol  oscillators}

A system of $M$ coupled Van der Pol oscillators may exhibit chaos without external driving. Such a system is described by:
\begin{eqnarray}\label{Pol}
\frac{\partial x_i}{\partial t} = y_i, \nonumber \\
\frac{\partial y_i}{\partial t} = \left [ \epsilon_i-\left (x_i+\sum_j {\kappa_{ij}x_j}\right )^2 \right ]y_i-
\left (x_i+\sum_j {\kappa_{ij}x_j}\right )
\end{eqnarray}
The parameter $\epsilon_i$ determines the limit cycle of oscillator $i$ (for $\kappa=0$), while $\kappa_{ij}$ specifies the non-linear coupling between oscillator $i$ and oscillator $j$~\cite{Pastor:1993}.

\begin{figure}\centering
  \includegraphics[trim=0 0 0 0,clip=,width=16cm]{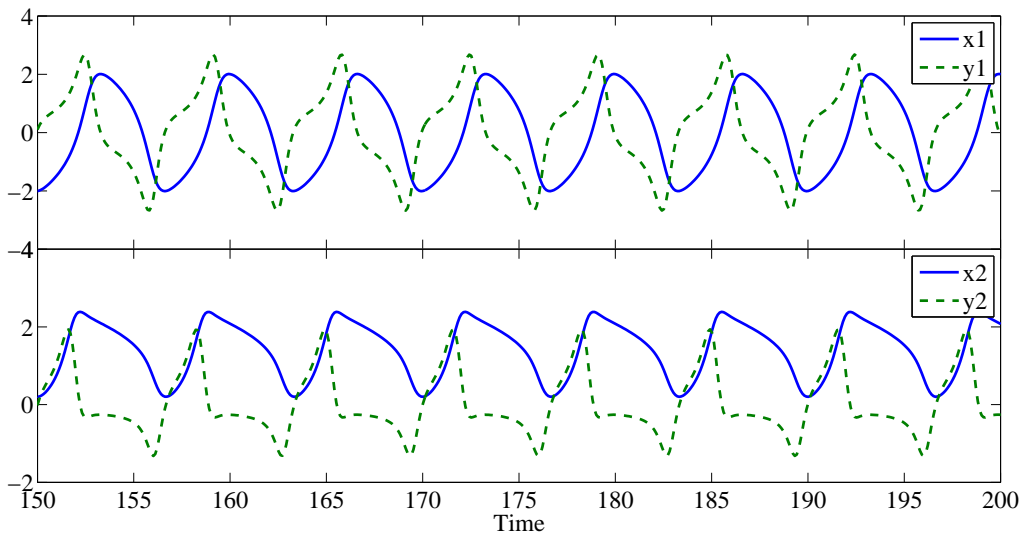}
\caption{\label{Pol_data}Time traces obtained from the system of $M=2$ Van der Pol oscillators.}
\end{figure}

We have run a simulation with $M=2$, $\epsilon =(1,1.1)$ and
\begin{equation}
\kappa = \left(\begin{array}{cc}0 & 1 \\0 & 0\end{array}\right),
\end{equation}
meaning that there are two oscillators with slightly different limit cycles, while oscillator 2 affects oscillator 1 (but not vice versa).
With this choice of parameters, the system is in a quasi-periodic state.
Time was integrated from $t=0$ to $t=1000$, and 10000 equally spaced data points were saved for analysis. 
Only the time interval from $t=100$ to $t=1000$ was used for analysis in order to remove the initial transient phase. 
A section of data is shown in Fig.~\ref{Pol_data}.

From a spectral analysis, the mean oscillation period of the signals was about 6.72, corresponding to about 67 samples per period.
Thus, $\alpha=\beta$ should be chosen less than about 17 ($\simeq 67/4$). The latter value (17) also roughly corresponds to the first minimum of the self-Mutual Information of $x_1$.

\begin{figure}\centering
  \includegraphics[trim=0 0 0 0,clip=,width=12cm]{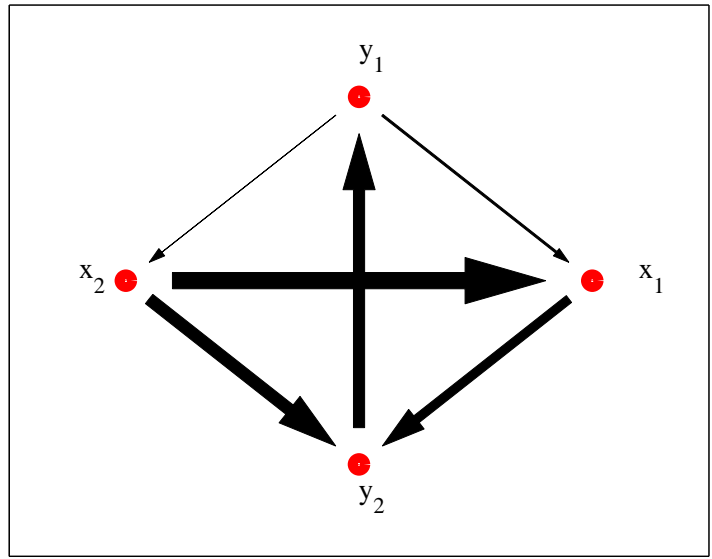}
\caption{\label{Pol_flow}Information flow between the four signals of the system of $M=2$ Van der Pol oscillators: the arrows indicate the direction of flow and the width is proportional to the amount of information transfer.}
\end{figure}

The selected data are analyzed with $\alpha=\beta=8$, $m=3$.
Net information flow from signal $i$ to signal $j$ is computed as $T^{\rm net}_{ij} = T_{ij}-T_{ji}$, where the indices $1, \dots, 4$ correspond to the signals $x_1,y_1,x_2,y_2$, respectively.
The following net Transfer Entropy matrix is obtained (only the part of the matrix above the diagonal is shown; the remainder follows from antisymmetry):
\begin{equation}
T^{\rm net} = \left(\begin{array}{cccc}
$0$ & $-0.036$ & $-0.29$ & $0.15$ \\
- & $0$ & $0.0007$ & $-0.20$ \\
- & - & $0$ & $0.21$\\
- & - & - & $0$
\end{array}\right)
\end{equation} 
This can be represented graphically by drawing 4 dots representing the 4 signals in a plane, cf. Fig.~\ref{Pol_flow}. The four dots are connected by arrows, such that the direction of the arrow indicates the direction of net information flow, while the width of the arrow is proportional to the value $T^{\rm net}_{ij}$.
There is a strong flow from $x_2$ to $x_1$ (corresponding to $T^{\rm net}_{13}$). 
This non-trivial component corresponds to the fact that $\kappa$ is such that oscillator 2 affects oscillator 1 (but not vice versa).
Another strong flow is from $y_2$ to $y_1$ ($T^{\rm net}_{24}$), for the same reason.
The flow from $x_2$ to $y_2$ ($T^{\rm net}_{34}$) is trivial (cf.~Eq.~(\ref{Pol})).
The remaining arrows are smaller, so that it is clear that the information flow is dominantly from oscillator 2 to oscillator 1.
Therefore, the analysis technique correctly recovers the direction of coupling among the components of the system.

\begin{figure}\centering
  \includegraphics[trim=0 0 0 0,clip=,width=12cm]{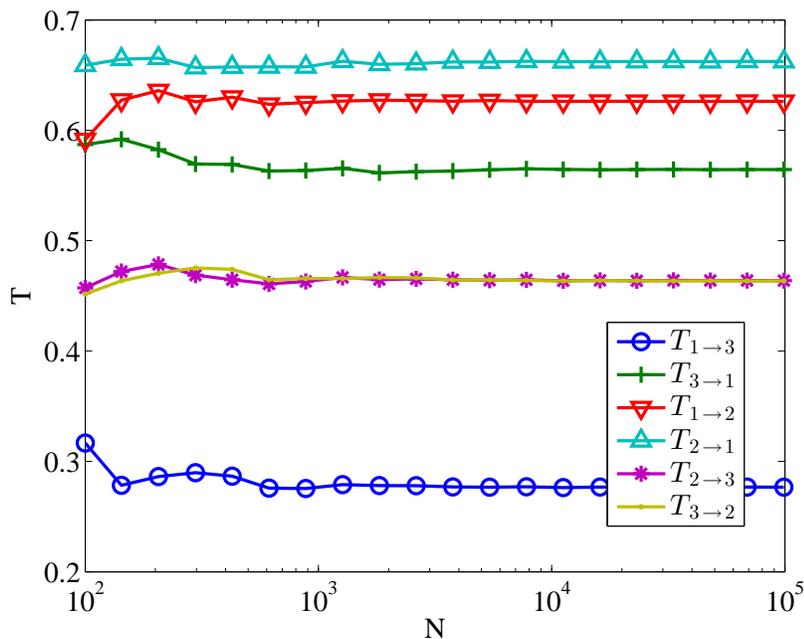}
\caption{\label{Pol_convergence}Numerical convergence of the Transfer Entropy versus length of the data arrays, $N$, for the system of $M=2$ coupled Van der Pol oscillators ($m=3$, $\alpha=\beta=8$).}
\end{figure}

The stability of the analysis method was tested by computing the Transfer Entropy for a varying length of the data arrays, $N$.
Fig.~\ref{Pol_convergence} shows $T_{13}$ (the Transfer Entropy from signal 1, $x_1$, to signal 3, $x_2$) and $T_{31}$ 
(the Transfer Entropy from signal 3, $x_2$, to signal 1, $x_1$) versus $N$, with the analysis settings as in the previous paragraph.
It is seen that $T$ converges to a stable value for $N\gtrsim 10^3$, a rather modest number. 
Note that in this case, the total number of bins of the main pdf is $m^{l+k+1}=3^3=27$.
The value of the Transfer Entropy $T$, expressed in bits, can be calibrated against the total bit range, $\log_2 m = 1.58$, implying that the coupling strength is quite significant.

\begin{figure}\centering
  \includegraphics[trim=0 0 0 0,clip=,width=12cm]{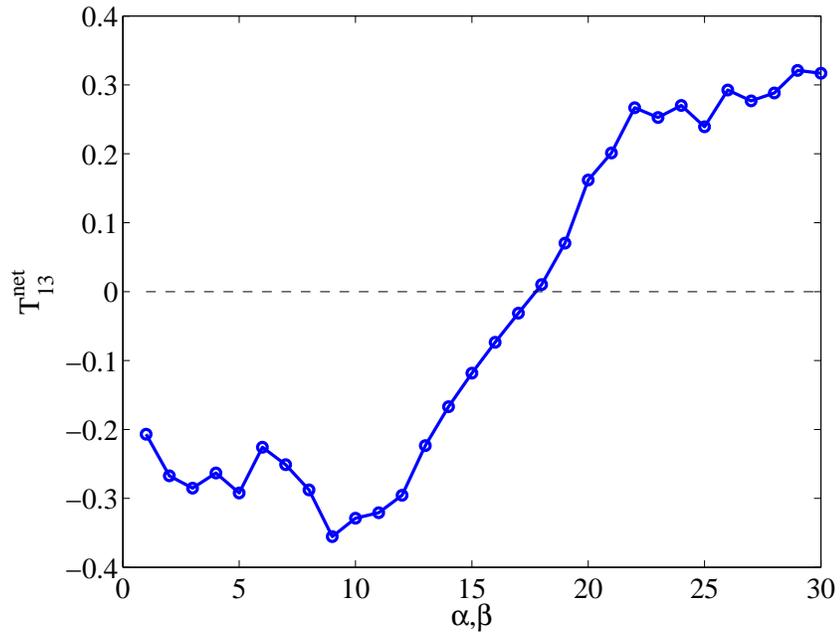}
\caption{\label{Pol_netflow}Net information flow between the two coordinates $x_1$ and $x_2$ of the system of $M=2$ Van der Pol oscillators (flow from $x_1$ to $x_2$ being positive), for different values of $\alpha=\beta$.}
\end{figure}

To understand the evolution of the Transfer Entropy with the analysis parameters $\alpha,\beta$, we
calculated $T^{\rm net}_{13}$ (which specifies the net flow from signal 1, $x_1$, to signal 3, $x_2$) for a range of $\alpha=\beta$ values, cf.~Fig.~\ref{Pol_netflow}.
The graph shows that the net information transfer from $x_1$ to $x_2$ is negative for $\alpha,\beta \le 17$ (as it should, for we know that the information transfer should go from oscillator 2 to oscillator 1). For higher values of $\alpha,\beta$, the net flow changes sign.
This occurs when crossing the quarter-period value (17) or minimum self-Mutual Information value (17), and is caused by the fact that this system is quasi-periodic.
It seems important, therefore, to keep $\alpha,\beta$ well below the mentioned reference values.

\clearpage
%===========================
\subsection{A simplified predator-prey model}

In the context of fusion plasmas, spontaneous confinement transitions are of prime interest.
In recent years, models have been developed to describe such transitions, involving the nonlinear interaction between various fields.
In this section, we will use the model of Ref.~\cite{Pilarczyk:2012} to generate signals for analysis using the Transfer Entropy technique.
The model equations are:
\begin{eqnarray}\label{PP_model}
\frac{dE}{dt} = \left ( \frac{1}{1+V'^2+U'^2} - E \right ) E \nonumber \\
\frac{dV'}{dt} = \left ( aE^2 + cU'^2 -b \right ) V' \\
\frac{dU'}{dt} = \left ( \frac{aE^2}{1+V'^2} - b \right ) U' + dE^2 V' \nonumber 
\end{eqnarray}
Here, $E$ represents the turbulence amplitude, $U'$ the Zonal Flow shear, and $V'$ the sheared flow.

\begin{figure}\centering
  \includegraphics[trim=0 0 0 0,clip=,width=16cm]{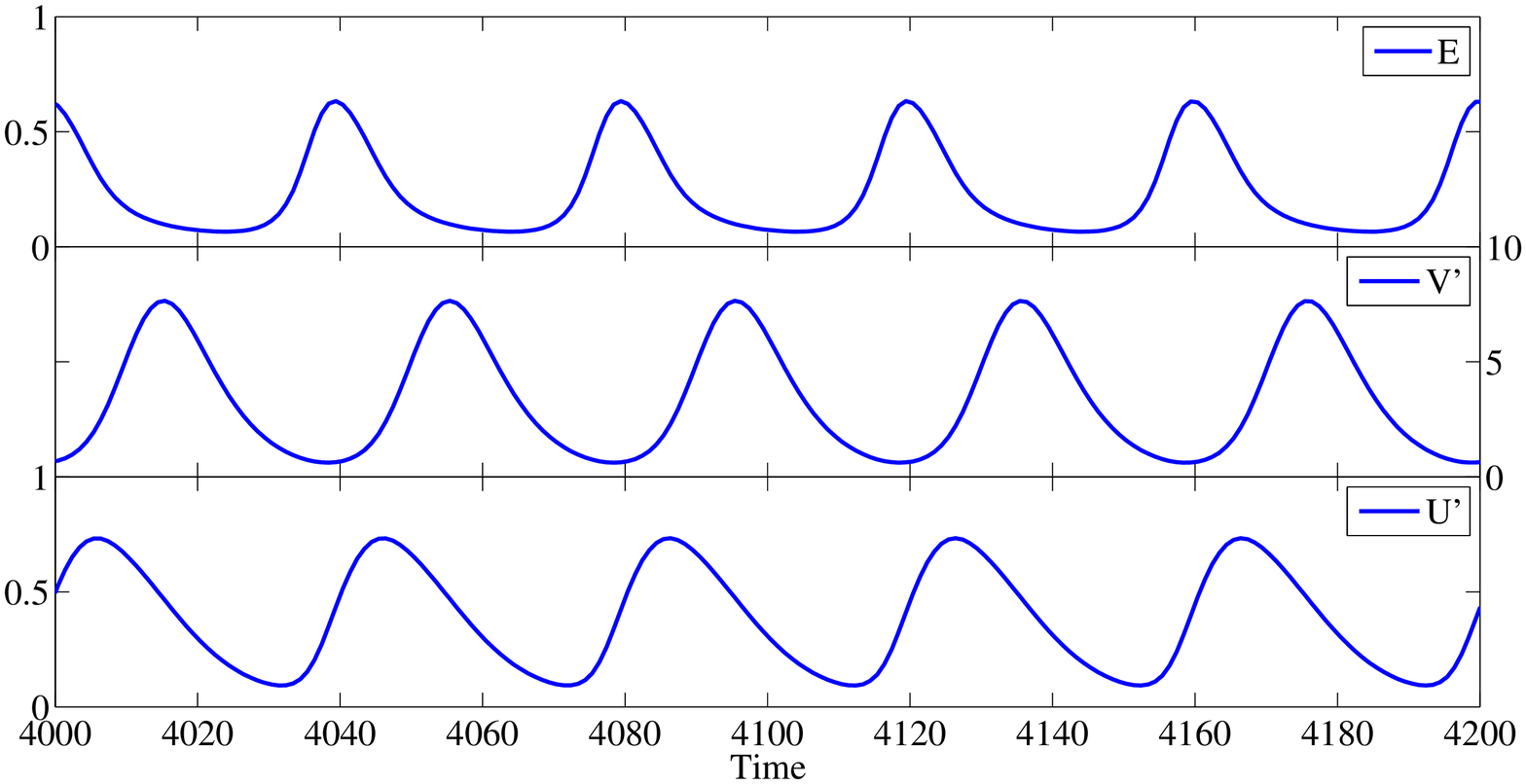}
\caption{\label{PP_signals}Signals from a run of the model of Eq.~(\ref{PP_model}) with parameters $a=0.204, b=0.16, c=0.714$ and $d=0.5$.}
\end{figure}

We performed a simulation run with $a=0.204, b=0.16, c=0.714$ and $d=0.5$ (cf.~Fig.~5 of the cited paper), generating a set of 10,000
time points (at sampling rate $\Delta t=1$). A short section of the model output is shown in Fig.~\ref{PP_signals}.
The mean period of the quasi-periodic oscillations is 40.09, corresponding to about 40 samples.
The first minimum of the self-Mutual Information occurs at 8 (for $U'$), 10 (for $V'$) or 14 (for $E$) samples.
Thus, $\alpha$ and $\beta$ should be chosen well below 8.

The Transfer Entropy is computed for all 9 possible combinations of signals ($1=E,2=V',3=U'$).
The settings chosen are: $\alpha = \beta = 5, m = 5$. The following Transfer Entropy matrix is obtained:
\begin{equation}
T = \left(\begin{array}{ccc}
0 & 0.4913 & 1.0791 \\
0.6320 & 0 & 0.8540 \\
0.7084 & 1.0809 & 0
\end{array}\right)
\end{equation}
Comparing the values of $T$ with $\log_2 m = 2.32$, one concludes that the interactions are quite strong.

\begin{figure}\centering
  \includegraphics[trim=0 0 0 0,clip=,width=12cm]{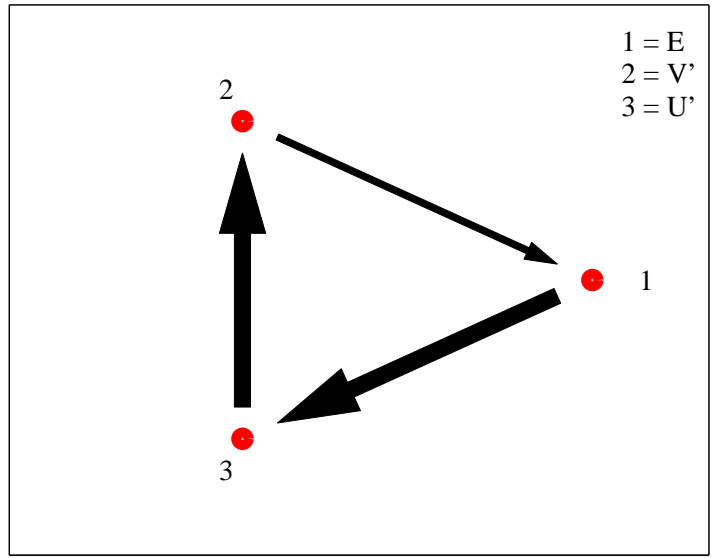}
\caption{\label{PP_flow}Information flow between the three signals of the predator-prey model: the arrows indicate the direction of flow and the width is proportional to the amount of information transfer.}
\end{figure}

Net information flow from signal $i$ to signal $j$ is computed as $T^{\rm net}_{ij} = T_{ij}-T_{ji}$. 
This is again represented graphically by drawing 3 dots representing the 3 signals in a plane. 
The three dots are connected by arrows, such that the direction of the arrow indicates the direction of net information flow, while the width of the arrow is proportional to the value $T^{\rm net}_{ij}$.
See Fig.~\ref{PP_flow}.

The resulting diagram makes eminent physical sense: 
$E$ drives $U'$ (turbulence drives Zonal Flow);
$U'$ drives $V'$ (Zonal Flow drives sheared flow);
$V'$ drives $E$ (sheared flow controls turbulence).
This is precisely the order of interaction that one would expect in this type of model.
Of course, in this simplified case this sequence can also be obtained by simple inspection of the data (Fig.~\ref{PP_signals}), as 
the mutual time delays are in clear accord with this result.
On the other hand, the present method is generally applicable and allows quantifying the result.

%\clearpage
%%===========================
%\subsection{A radial transport model}
%
%{[Pending?] Model by D.E. Newman et al., cf. \href{http://ocs.ciemat.es/EPS2013PAP/pdf/P1.173.pdf}{EPS 2013 - P1.173}.}

\clearpage
%===========================
\section{Experimental data analysis results}\label{results}

In this section, we will apply the Transfer Entropy technique to data obtained from various magnetic confinement devices.
The data have been selected for their relevance to the study of the interaction between Zonal Flows and turbulence.
Zonal Flows are large scale electrostatic potential structures that form spontaneously in magnetically confined toroidal fusion plasmas, and have zero toroidal wavenumber, small or zero poloidal wavenumber and finite radial wavenumber~\cite{Diamond:2005}.
The global nature of these structures makes them hard to identify, as most measurements (of potential or radial electric field) are local.
In the following, we will not worry about the precise identification of Zonal Flows, but confide in earlier published analyses showing that the presented data pertain to Zonal Flows with high probability.
Our main goal here is to analyze the interaction between these hypothetical Zonal Flows (identified via potential or radial electric field fluctuations) and turbulence (identified via the density fluctuation amplitude or radial particle transport flux).

\subsection{TJ-K}

\begin{figure}\centering
  \includegraphics[trim=0 0 0 0,clip=,width=16cm]{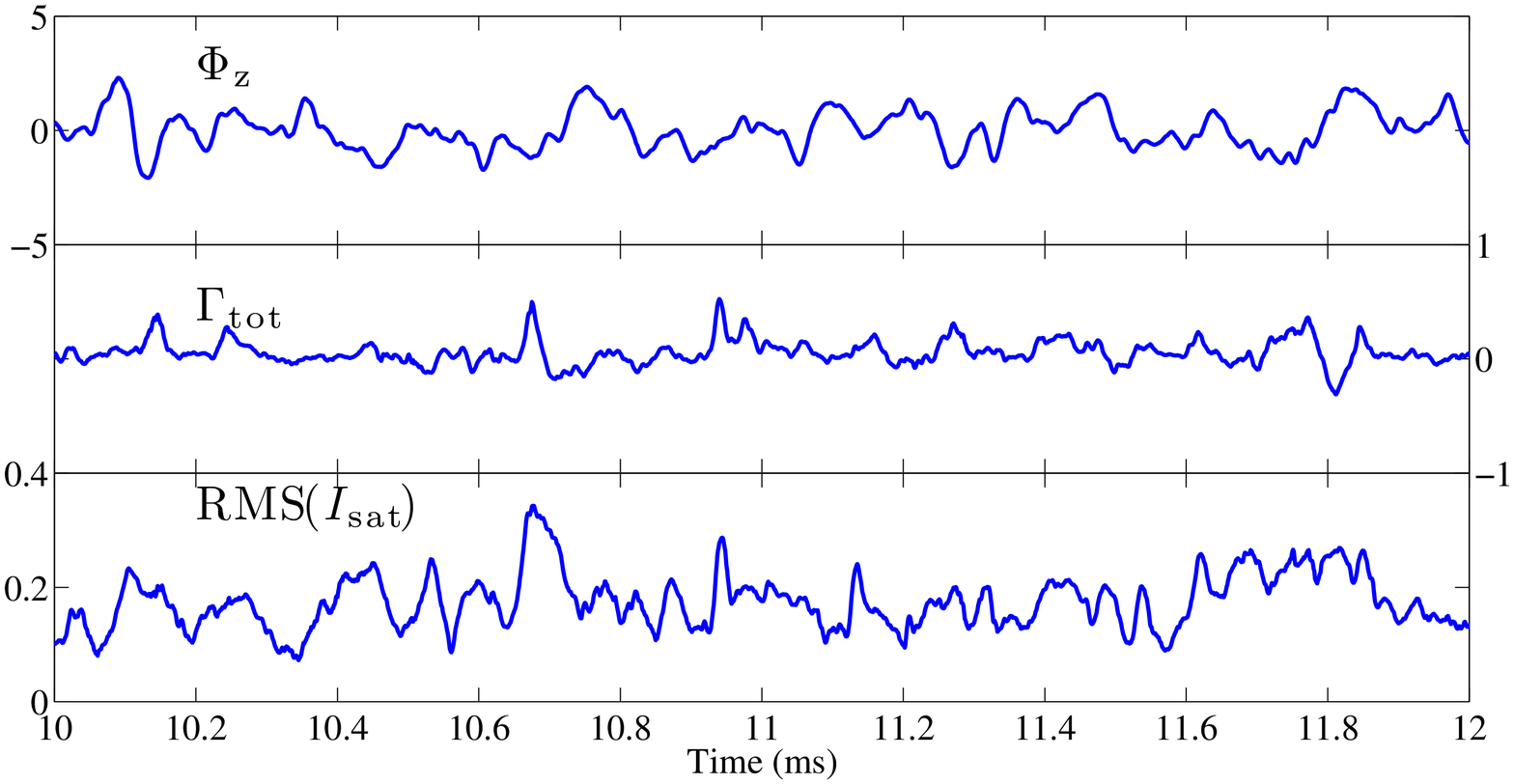}
\caption{\label{TJK_data}A short section of data from TJ-K (arbitrary units).}
\end{figure}

Here, we analyze data from the TJ-K stellarator, a torsatron operated at low magnetic field ($B = 72$ mT) and low plasma beta.
The discharge analyzed here corresponds to a helium plasma, heated by microwaves, with a central density of $n = 2.3 \cdot 10^{17}$ m$^{-3}$ and an electron temperature of $T_e = 8$ eV and cold ions, as reported in more detail elsewhere~\cite{Birkenmeier:2013b}.
In this experiment, turbulence was dominated by electrostatic drift wave turbulence, and the total particle transport and zonal potential were found to be linked in a predator-prey cycle.
Among other diagnostics, the device disposes of a set of 64 Langmuir probes, distributed over a poloidal circumference of the device.
The probes are configured to measure floating potential and ion saturation current in an alternating fashion, at a sampling rate of 1 MHz.
From these signals, we compute the zonal potential $\Phi_z(t)$ as the mean poloidal value of the floating potential, and the global radial particle flux $\Gamma_{\rm tot}(t)$ as the poloidal mean of the local radial particle flux, proportional to the fluctuating ion saturation current times the local poloidal electric field, as described in more detail in the cited reference.
We quantify the `global turbulence level' by computing the root mean square (RMS) deviation of the 32 poloidally distributed ion saturation current measurements ($I_{\rm sat}$), thus obtaining a quantifier of the turbulence level with the same time resolution as $\Phi_z(t)$ and $\Gamma_{\rm tot}(t)$. 
A short section of data is shown in Fig.~\ref{TJK_data}; clearly, these data are much less regular than the model data shown in the preceding section, making it considerably more difficult to understand the nonlinear relationship between the signals.

\begin{figure}\centering
  \includegraphics[trim=0 0 0 0,clip=,width=12cm]{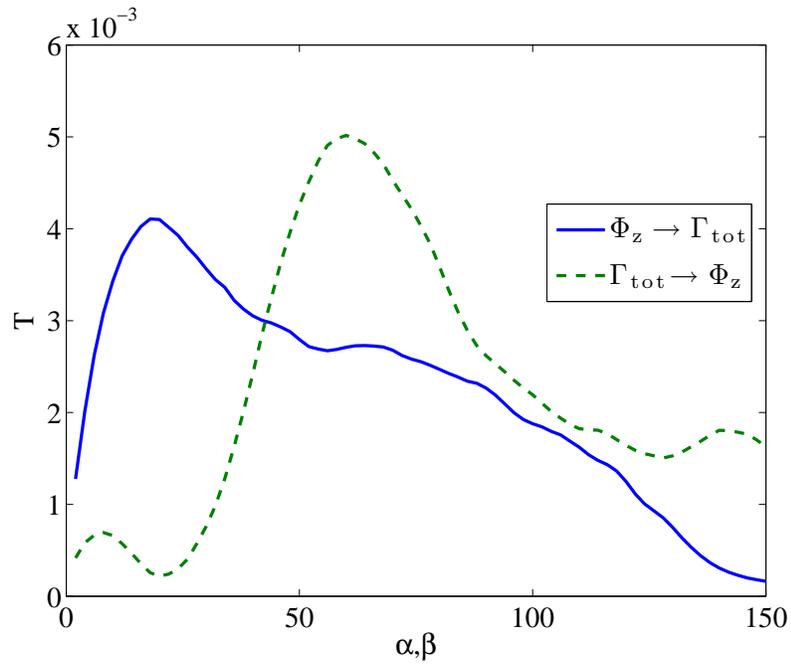}
\caption{\label{TJK}Transfer Entropy between the zonal potential $\Phi_z(t)$ and the global particle transport $\Gamma_{\rm tot}(t)$ at TJ-K versus $\alpha=\beta$ (in sampling units, i.e., $\mu$s).}
\end{figure}

We apply the analysis described above, setting $m=3$ (coarse graining). 
Fig.~\ref{TJK} shows the Transfer Entropy between the two signals $\Phi_z(t)$ and $\Gamma_{\rm tot}(t)$ as a function of $\alpha = \beta$.
The amplitude of the Transfer Entropy is rather small, namely below $5\cdot 10^{-3}$, compared to the full bit range $\log_2 m = 1.59$, which indicates that the causal link between these variables is not very strong.
Nevertheless, it is unexpected and interesting to observe that $T_{\Phi \to \Gamma}$ and $T_{\Gamma\to \Phi}$ peak at different values of $\alpha = \beta$.
$T_{\Phi \to \Gamma}$ peaks at about 20 $\mu$s, while
$T_{\Gamma\to \Phi}$ peaks at about 60 $\mu$s.
Thus, zonal potential $\Phi_z$ has a rather fast impact on the total particle flux $\Gamma_{\rm tot}$, while the
total particle flux $\Gamma_{\rm tot}$ acts back on the zonal potential $\Phi_z$ on a much longer time scale.
In terms of net information transfer, it flows from $\Phi_z$ to $\Gamma_{\rm tot}$ for time scales less than about 40 $\mu$s, and in the opposite direction for longer time scales.
The two distinct time scales for mutual interaction would immediately give rise to oscillatory behavior, as indeed observed.
This seems coherent with the usual predator-prey models for the interaction between Zonal Flow and turbulence.

\begin{figure}\centering
  \includegraphics[trim=0 0 0 0,clip=,width=12cm]{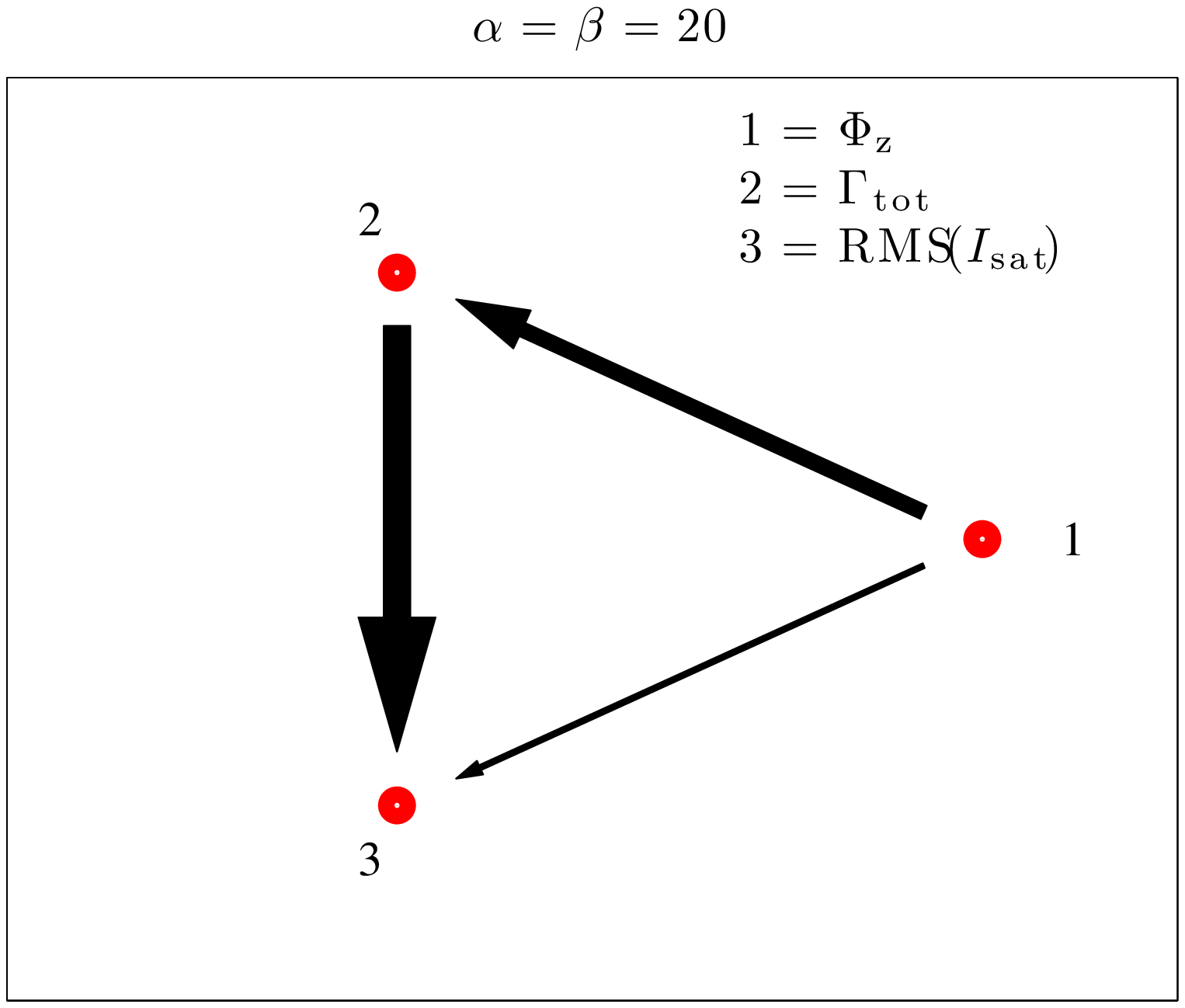}\\
  \includegraphics[trim=0 0 0 0,clip=,width=12cm]{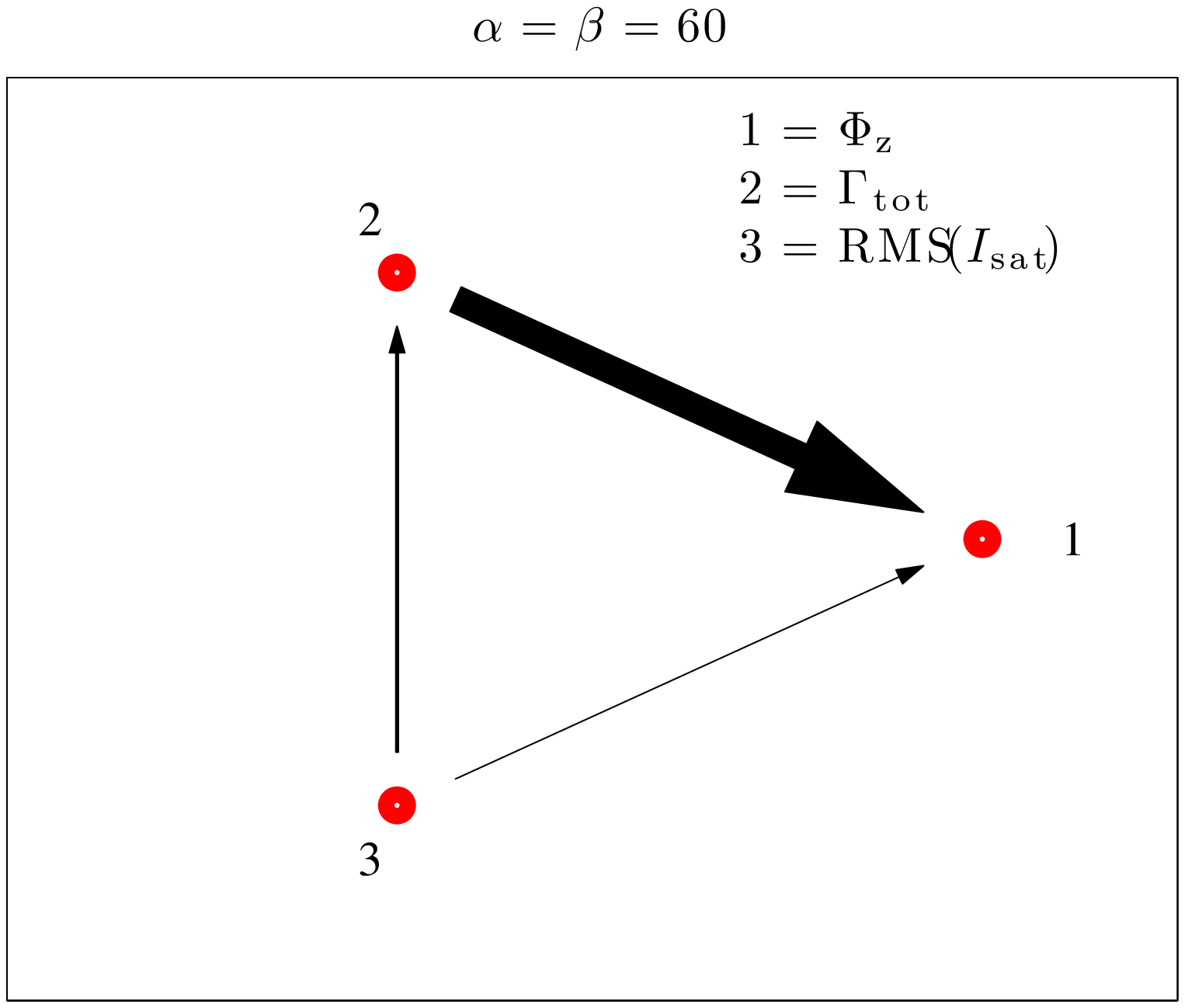}
\caption{\label{TJK_cycle}Graphical representation of the net Transfer Entropy between the zonal potential $\Phi_z(t)$, the global particle transport $\Gamma_{\rm tot}(t)$, and the global turbulence level RMS$(I_{\rm sat}$) at TJ-K for two values of $\alpha=\beta$.}
\end{figure}

In the standard Zonal Flow model, the Zonal Flow has an impact on the global turbulence level.
Fig.~\ref{TJK_cycle} shows the interaction diagram between all three signals at the two most significant values of $\alpha=\beta$.
On a short time scale (20 $\mu$s), the Zonal Flow affects the transport, and the transport in turn affects the turbulence level (presumably, by modifying the driving gradients).
On a longer time scale (60 $\mu$s), the transport affects the Zonal Flow, but the interaction with the turbulence level is insignificant.

The short time scale result is interesting, as it confirms the analysis of~\cite{Birkenmeier:2013b}, where it was observed that the Zonal Flow does not affect the turbulence amplitude strongly, but rather it affects the transport (as noted in the cited paper, by modifying the phase relation between density and potential fluctuations). The modification of the transport then affects the turbulence amplitude.
Although there is an arrow showing that the zonal potential also affects the turbulence level directly, its strength is much less than the indirect route via the turbulent transport.
The long time scale result is presumably simply due to a restoration of ambipolarity: a modification of transport must eventually lead to a modification of potential. 
%{[Birkenmeier: could this be Stringer spinup? Stringer T 1969 Phys. Rev. Lett. 22 770]}

\clearpage
%===========================
\subsection{TJ-II: Doppler reflectometry}

TJ-II is a Heliac type stellarator with 4 field periods. 
The experiments discussed below have been carried out in pure Neutral Beam Injection (NBI) heated plasmas (line averaged electron density $\langle n_e \rangle = 2-4 \times 10^{19}$ m$^{-3}$, central electron temperature $T_e = 300-400$ eV, {$T_i \simeq 140$ eV}).
The input NBI power was about 500 kW.
These discharges have been reported elsewhere in more detail~\cite{Estrada:2011,Happel:2011,Milligen:2013}.

In this section, we will analyze data from the Doppler reflectometry diagnostic taken as the plasma experiences spontaneous confinement transitions.  
In Doppler reflectometry, a finite tilt angle is purposely introduced between the incident probing beam and the normal to the reflecting cut-off layer, and the Bragg back-scattered signal is measured~\cite{Hirsch:2001}. 
The amplitude of the recorded signal, $A$, is a measure of the intensity of the density fluctuations, $\tilde n$. 
Furthermore, as the plasma rotates in the reflecting plane (flux surface), the scattered signal experiences a Doppler shift.
The size of this shift is directly proportional to the rotation velocity of the plasma turbulence perpendicular to the magnetic field lines, $v_\perp$, and therefore to the plasma background $E \times B$ velocity, provided the latter dominates over the phase velocity of density fluctuations (cf.~\cite{Estrada:2009}). 
The Doppler reflectometer signals, sampled at 10 MHz, allow determining $\tilde n$ and $v_\perp$ with high temporal resolution.

\begin{figure}\centering
  \includegraphics[trim=0 0 0 0,clip=,width=12cm]{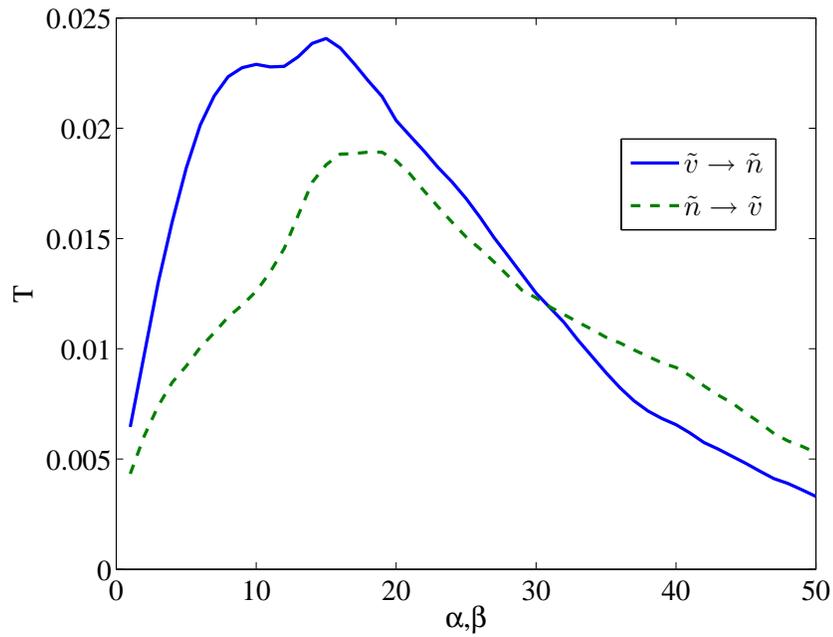}
\caption{\label{30895_ab}Transfer Entropy between $\tilde v_\perp$ (the fluctuating perpendicular flow velocity) and $\tilde n$ (the turbulence amplitude) in the I-phase versus $\alpha=\beta$ (in sampling units, {0.1 $\mu$s}).}
\end{figure}

First, we consider discharges in a magnetic configuration with edge rotational transform $\iota(a)/2\pi=1.553$. In this configuration, a transition from L-mode to an Intermediate (I) phase is often observed (intermediate between the L and H modes).
In the I-phase, predator-prey oscillations occur, and bicoherence is relatively strong as reported elsewhere~\cite{Milligen:2013}.
Fig.~\ref{30895_ab} shows an example of the Transfer Entropy for data in a 20 ms long time window in the I-phase versus $\alpha=\beta$ (with $m=3$).
The graph bears similarity to the corresponding graph for TJ-K, Fig.~\ref{TJK}, in that there is a clear peak in the Transfer Entropy curves, while $T_{\tilde v \to \tilde n}$ dominates over $T_{\tilde n \to \tilde v}$ for small values of $\alpha,\beta$.
The position of the peak of the Transfer Entropy appears to be related to the autocorrelation time of the turbulence ($\sim 50~\mu$s for TJ-K, $1-10~\mu$s for TJ-II).
Thus, it is not related to the very slow predator prey cycles reported in earlier work~\cite{Estrada:2011}, with a period of about a ms.
In other words, the analysis based on the Transfer Entropy has uncovered a novel interaction.
%{[Note, perhaps interesting: the I-phase data have secondary $T$ peaks for larger values of $\alpha,\beta$, which is not the case for the H phase data. The problem is that these secondary peaks have low amplitude.]}

\begin{figure}\centering
  \includegraphics[trim=0 0 0 0,clip=,width=12cm]{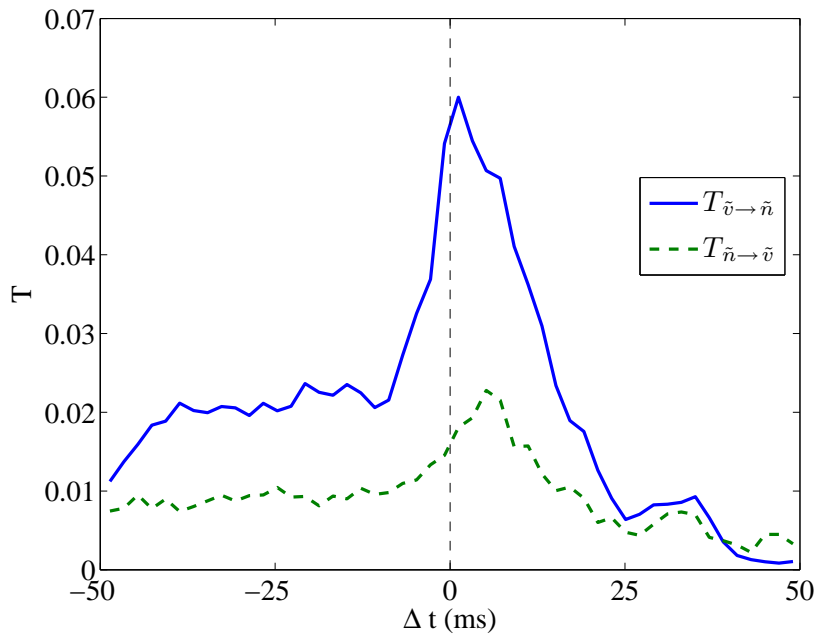}
\caption{\label{Series_100_35}Mean Transfer Entropy between $\tilde v_\perp$ (the fluctuating perpendicular flow velocity) and $\tilde n$ (the turbulence amplitude) for 10 {discharges} in a magnetic configuration with $\iota(a)/2\pi=1.553$ (the L--I transition occurs at $\Delta t = 0$).}
\end{figure}

%In discharge 27123, this transition occurs around $t = 1168$ ms.
%For $t>1168$ ms, predator-prey oscillations occur, and bicoherence is relatively strong as reported elsewhere~\cite{Milligen:2013}.
%The analysis of Transfer Entropy between the two signals $1 = \tilde v_\perp$ (the fluctuating perpendicular flow velocity) and $2 = \tilde n$ (the turbulence amplitude) measured by the Doppler reflectometer yields the graph shown in Fig.~\ref{27123}.
%The Transfer Entropy is computed for successive 2 ms time sections using $m=3$, $\alpha=\beta=10$.
%Prior to the transition, in L mode, $T_{1\to 2}$ and $T_{2\to 1}$ are roughly balanced and relatively small.
%However, after the transition, in the I mode characterized by predator-prey oscillations, $T_{1\to 2}>T_{2\to 1}$, indicating there is a net information flow from $\tilde v_\perp$ to $\tilde n$ or, in other words, the (zonal) flow is regulating the turbulence.
%\begin{figure}\centering
%  \includegraphics[trim=0 0 0 0,clip=,width=12cm]{27123}
%\caption{\label{27123}Transfer Entropy between the two signals $1 = \tilde v_\perp$ (the fluctuating perpendicular flow velocity) and $2 = \tilde n$ (the turbulence amplitude) measured by the Doppler reflectometer for discharge 27123. An L--I transition occurs at $t=1168$ ms.}
%\end{figure}
%
Fig.~\ref{Series_100_35} shows the {mean} evolution of the Transfer Entropy for 10 {discharges} in this magnetic configuration. 
{ In these discharges, an L--I transition occurred at a certain time, which was defined as $\Delta t=0$ ms.
The time window $-50 \le \Delta t \le 50$ ms was subjected to analysis.
In this time window, the Transfer Entropy was computed for successive 2 ms time sections of the signals $\tilde v$ and $\tilde n$, using $m=3$, $\alpha=\beta=10$. Finally, the resulting Transfer Entropy curves were averaged over the 10 selected discharges.
}

%
%Fig.~\ref{23048} shows a similar graph, for discharge 23048, configuration 100\_44, in which an L--H transition occurs at $t=1137$ ms.
%Here, the phase of net information flow (where $T_{1\to 2}>T_{2\to 1}$) starts clearly before the transition, and decays afterward.
%\begin{figure}\centering
%  \includegraphics[trim=0 0 0 0,clip=,width=12cm]{23048}
%\caption{\label{23048}Transfer Entropy between the two signals $1 = \tilde v_\perp$ (the fluctuating perpendicular flow velocity) and $2 = \tilde n$ (the turbulence amplitude) measured by the Doppler reflectometer for discharge 23048. An L--H transition occurs at $t=1137$ ms.}
%\end{figure}

\begin{figure}\centering
  \includegraphics[trim=0 0 0 0,clip=,width=12cm]{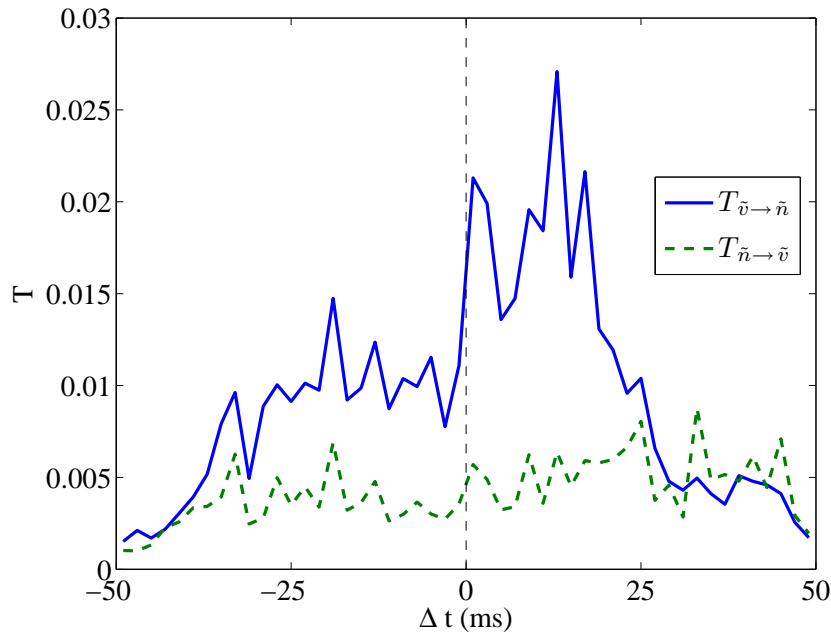}
\caption{\label{Series_101_42}Mean Transfer Entropy between $\tilde v_\perp$ (the fluctuating perpendicular flow velocity) and $\tilde n$ (the turbulence amplitude) for 4 {discharges} in a magnetic configuration with $\iota(a)/2\pi=1.630$ (the L--H transition occurs at $\Delta t = 0$).}
\end{figure}

Next, we consider discharges in a magnetic configuration with $\iota(a)/2\pi=1.630$. In this configuration, a relatively rapid transition from L-mode to H-mode is often observed, without intermediate (I) phase~\cite{Happel:2011}.
{The average Transfer Entropy was computed for a number of discharges using an analogous procedure as described above, however setting $\Delta t=0$ at the L--H transition time.}
Fig.~\ref{Series_101_42} shows the average evolution of the Transfer Entropy for 4 {discharges} in this magnetic configuration (around the L--H transition).
The Transfer Entropy $T_{\tilde v \to \tilde n}$ increases sharply by a factor of 2 at the L--H transition, indicating the regulation of turbulence ($\tilde n$) by the Zonal Flow ($\tilde v$). This regulatory phase lasts for about $15-20$ ms, in accord with the duration of enhanced bicoherence reported elsewhere~\cite{Milligen:2013}.

We draw attention to an interesting difference between the L--I and L--H transitions. 
With the L--H transition, the transition is followed by a rapid increase in $T_{\tilde v \to \tilde n}$, while $T_{\tilde n \to \tilde v}$ remains approximately constant.
Thus, the Zonal Flow is simply regulating the turbulence (suppressing it).
With the L--I transition, the transition also shows a rapid increase of $T_{\tilde v \to \tilde n}$, but this is mirrored (although at a lower intensity level) by a similar increase in $T_{\tilde n \to \tilde v}$.
This is consistent with the fact that not only does the Zonal Flow regulate the turbulence, but the turbulence also acts back on the Zonal Flow, which could be related to the observed (predator-prey type) oscillations.
Also, in the case of the L--I transition, the values achieved by the Transfer Entropy are about 3 times higher than with the L--H transition.
In both cases, the amplitude of the Transfer Entropy is modest compared to the bit range, $\log_2 m = 1.59$, although an order of magnitude above the TJ-K case reported in the preceding section.

\clearpage
%===========================
\subsection{TJ-II: Langmuir probes}

\begin{figure}\centering
  \includegraphics[trim=0 0 0 0,clip=,width=12cm]{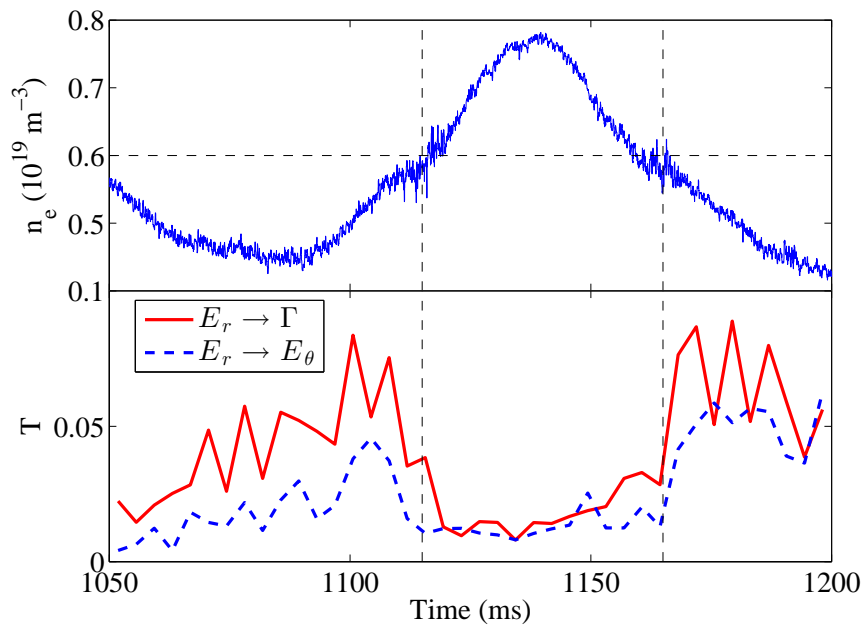}
\caption{\label{18080}{Top: Line averaged electron density for discharge 18080. The horizontal dashed line indicates the approximate value of the critical density; vertical dashed lines indicate the approximate limits of the enhanced confinement state ($1115 < t < 1165$ ms).
Bottom: Transfer Entropy between various relevant derived Langmuir probe signals for discharge 18080. The Langmuir probe is located at radial position $\rho \simeq 0.92$. The net Transfer Entropy $T^{\rm net}$ for these signal combinations is very similar.}}
\end{figure}

In discharge 18080, heated by Electron Cyclotron Resonant Heating ($P_{\rm ECRH} \simeq 400$ kW), a triple Langmuir probe was inserted to normalized radius $\rho = 0.92$.
By raising the electron density, a spontaneous confinement transition was provoked, and a subsequent back-transition was achieved by bringing the density down again~\cite{Milligen:2008c}. 
It should be noted that this transition is not an L--H transition, but is related to a change of Neoclassical root~\cite{Velasco:2012} (a local  sign change of the mean radial electric field, $\overline{E_r}$).
The density evolution is shown in {Fig.~\ref{18080} (top)}, showing the double crossing of the critical line averaged density value.

The Langmuir probe measured floating potentials and ion saturation currents on various pins at a sampling rate of 1 MHz.
The probe configuration allowed the computation of the fluctuating radial and poloidal electric fields, $E_r$ and $E_\theta$, and the fluctuating radial particle flux $\Gamma$.
{Fig.~\ref{18080} (bottom)} shows the Transfer Entropy between some of these signals, computed for successive 2 ms time sections using $m=3$, $\alpha=\beta=5$.

Interestingly, the Transfer Entropy is largest for the combination $E_r \to \Gamma$. 
This is significant, in view of the fact that this corresponds to the impact of a possible Zonal Flow ($E_r \propto v_\theta$) on the radial particle flux.
This quantity is seen to build up gradually before the transition, and essentially disappear during the enhanced confinement state ($1115 < t < 1165$ ms). The build-up phase presumably corresponds to the gradual development and growth of a Zonal Flow, 
which however disappears when the line averaged density is above its critical value, $n_{\rm crit} \simeq 0.6 \cdot 10^{19}$ m$^{-3}$. 
The Transfer Entropy is also large for the combination $E_r \to E_\theta$.
This is also significant, as Sheared Flow is produced by Reynolds Stress according to standard Zonal Flow models, which can only be large if $E_r$ and $E_\theta$ are phase-correlated.
Traditional analyses have indeed shown that this phase correlation occurs~\cite{Alonso:2009}, but the present analysis adds the information  that it is the Zonal Flow ($E_r$ or poloidal velocity) that drives the poloidal electric field (or radial particle velocity), and not the other way around. 
After the back-transition, all quantities return approximately to their pre-transition values.
It is noted that very similar results are obtained for a set of 6 similar discharges, showing that these results are robust.

\begin{figure}\centering
  \includegraphics[trim=0 0 0 0,clip=,width=12cm]{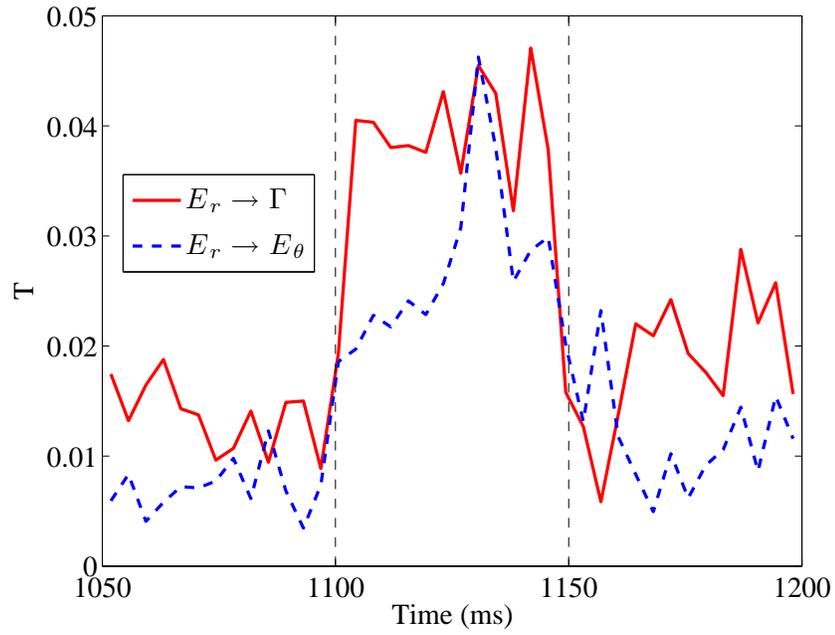}
\caption{\label{16014_16015}Mean Transfer Entropy between various relevant derived Langmuir probe signals for discharges 16014 and 16015. The Langmuir probe is located at radial position $\rho \simeq 0.81$. External biasing was applied between $t=1100$ and $t=1150$ ms. The net Transfer Entropy $T^{\rm net}$ for these signal combinations is very similar.}
\end{figure}

A similar analysis was made for two discharges with initial subcritical density in which external biasing was applied between $t=1100$ and $t=1150$ ms.
A biasing probe was inserted about 2 cm into the plasma and biased with respect to a poloidal limiter tangent to the last closed flux surface. The triple Langmuir probe was inserted to normalized radius $\rho \simeq 0.81$.
Detailed information about these discharges can be found elsewhere~\cite{Milligen:2008c}.
When applying positive biasing, turbulence was suppressed, leading to an improvement of confinement such that the density rose to values exceeding the critical density for spontaneous transitions ($\overline{n_e} > n_{\rm crit}$);
however, contrary to the spontaneous confinement transition, here $\overline{E_r}$ remains positive.
{Fig.~\ref{16014_16015}} shows the evolution of $T_{E_r\to E_\theta}$ and $T_{E_r \to \Gamma}$.
It is clear that biasing has a strong effect on these quantities.

Comparing the spontaneous and biasing-induced confinement transitions, one observes that $T_{E_r \to \Gamma}$ is large for
$\overline{n_e} \lesssim n_{\rm crit}$ with the spontaneous transition, 
while it is large for $\overline{n_e} > n_{\rm crit}$ with the induced transition.
The explanation for this apparent contradiction is related to the evolution of the mean electric field profile $\overline{E_r(\rho)}$, and will be addressed in Section \ref{discussion}.

\clearpage
%===========================
\section{Discussion and conclusions}\label{discussion}

\subsection{General considerations}

The analysis of the causal relation between fluctuating variables is of prime interest when studying complex nonlinear systems, and fundamental to reach a full understanding of such systems and develop realistic models. 
In this work, we use the concept of `causality' in the restricted sense referred to in Section \ref{introduction} (Wiener's `quantifiable causality').

{The Transfer Entropy technique~\cite{Schreiber:2000} allows} detecting a causal relation between variables that does not require very lengthy time series (although stationary state is still a requirement) and that does not rely on the assumption of weak turbulence. 
In essence, the analysis is based on the observation of a (significant) number of repetitive event sequences occurring in a pair of time series, which however may occur in an irregular manner.

As is the case with all methods for causality detection, an important caveat is due.
The method only detects the information transfer between measured variables.
If the net information flow suggests a causal link between two such variables, this may either be due to a direct cause/effect relation (in the restricted sense referred to above), or due to the presence of a third, undetected variable that affects both (e.g., with different delays, thus generating an {\it apparent} causal relation).
Thus, physical insight into the system is always needed to determine whether all relevant variables are being measured and to decide whether the net information flow actually corresponds to a causal link.

\subsection{Tests on numerical models}

When computing the Transfer Entropy for numerical data generated by multivariate nonlinear models, it was found that the direction of interaction between system variables could be recovered. 
E.g., in the system of two coupled Van der Pol oscillators, it was clearly established that oscillator 2 affected oscillator 1, but not vice versa, in accordance with the design of the system.
A similar statement can be made for the simplified predator-prey model.
Numerical convergence of the analysis was tested and guidelines for an efficient choice of analysis parameters ($m$, $\alpha$, and $\beta$) are provided.

\subsection{Application to experimental data and interpretation of results}

We have explored the application of this technique to some data from turbulent fusion plasmas.
The selected measurement data are relevant to the understanding of the important confinement transition, in which turbulence spontaneously generates a more ordered plasma state with reduced radial transport.

The analysis of the global potential and flux at TJ-K revealed the existence of two time scales: 20 and 60 $\mu$s.
On the short time scale, the Zonal Flow potential was shown to affect transport, which in turn affected turbulence.
On the longer time scale, the transport affected the Zonal Flow potential, which was hypothesized to be due to a restoration of ambipolarity.
The short time scale result is in accordance with the previous analysis of ~\cite{Birkenmeier:2013b}, whereas the longer time scale result is novel and reveals the potential significance of the technique to uncover new relationships.

Doppler reflectometry data from TJ-II taken across L--I and L--H transitions in NBI heated plasmas showed how the fluctuating perpendicular flow velocity $\tilde v$, again associated with Zonal Flows, affects the turbulence.
Across the L--H transition, $T_{\tilde v \to \tilde n}$ was found to increase sharply, while the reverse interaction $T_{\tilde n \to \tilde v}$ remained fairly constant.
However, the L--I transition was characterized by an increase in {\it both} these quantities ($T_{\tilde v \to \tilde n}$ being dominant).
The I-phase is characterized by quasi-periodic predator-prey oscillations~\cite{Estrada:2010c}, which however occur on a rather slow time scale (of the order of a ms) compared to the interactions found here, occurring on a $\mu$s time scale.
Of course, if the predator-prey oscillations affect the turbulence, as one assumes must be the case, then this effect must be detectable on this fast time scale. This seems to be what the Transfer Entropy succeeds in doing.
Note that the interaction between the slow time scale of the predator-prey cycle and the fast time scale of the turbulence autocorrelation time was hinted at already in a previous analysis based on the bicoherence~\cite{Milligen:2013}.

Langmuir probe data from TJ-II taken across a low-density confinement transition in an ECR heated plasma show how the Transfer Entropy between the fluctuating radial electric field (associated with Zonal Flows) and the particle flux, $T_{E_r \to \Gamma}$, gradually grows prior to the transition and essentially disappears once the transition has taken place.
By contrast, during bias-induced transitions in ECR heated plasmas, $T_{E_r \to \Gamma}$ increases sharply while biasing is applied.

The observed behavior of the Transfer Entropy in ECR heated plasmas can perhaps be understood as follows.
In the case of the spontaneous transition in ECR heated plasmas, $T_{E_r \to \Gamma}$ increases gradually as the density is raised towards the critical value, which is interpreted as a gradual growth of the Zonal Flow amplitude.
Simultaneously, $T_{E_r \to E_\theta}$ increases, which is interpreted as the build-up of Reynolds Stress, expected to produce a (steady state) Sheared Flow.
However, as the density is raised slowly, the plasma adjust the profiles in an attempt to maintain the ambipolarity condition.
At a certain point, the electron root solution of the ambipolarity equation disappears. 
Immediately prior to this point, the flow susceptibility is large (i.e., small changes in the ambipolar flux are associated with large changes in $E_r$), which can be interpreted in terms of a low Neoclassical viscosity~\cite{Velasco:2012}, leading to large amplitude Zonal Flows, consistent with the observed gradual growth of $T_{E_r \to \Gamma}$ as the critical density is approached from below.
Following the transition to the ion root state, flow susceptibility is suddenly strongly reduced (Neoclassical viscosity is high), so that zonal flows are strongly damped, which is consistent with the disappearance of both $T_{E_r \to \Gamma}$ and $T_{E_r \to E_\theta}$ as the critical density is crossed.

In the case of the biasing discharges, $T_{E_r \to \Gamma}$ and $T_{E_r \to E_\theta}$ increase sharply when the biasing is activated.
From previous work it is known that the externally applied radial electric field gives rise to Long Range Correlations~\cite{Pedrosa:2008}, which is consistent with the formation of a Zonal Flow, associated with the observed growth of $T_{E_r \to \Gamma}$ and $T_{E_r \to E_\theta}$. 
A possible explanation may be that the imposed electric field enhances the ambipolar electric field value (while remaining in the electron root state) and correspondingly enhances the flow susceptibility, leading to Zonal Flow enhancement.

%By contrast, during the L--H and L--I transitions in NBI heated plasmas, $T_{\tilde v \to \tilde n}$ rises sharply as the transition occurs and only decays gradually afterwards; here, therefore, $T_{\tilde v \to \tilde n}$  seems to measure an ongoing interaction (turbulence suppression by Zonal Flow) that persists even after the transition; no Neoclassical changes of root take place.

Comparing the TJ-K and TJ-II results, we note that the amplitude of the Transfer Entropy is about an order of magnitude higher in the latter device. Presumably, this corresponds to a larger Zonal Flow amplitude generated by a stronger drive (steeper gradients).

From the numerical and experimental examples examined in this work, it is concluded that the Transfer Entropy constitutes a powerful tool to unravel the causal relationship between nonlinearly interacting fields in complex systems.
It is expected that this technique may find applications in many fields of research.
 
%===========================
\section*{Acknowledgements}
Research sponsored in part by the Ministerio de Econom\'ia y Competitividad of Spain under project Nr.~ENE2012-30832.
Fruitful discussions with J.L.~Velasco are gratefully acknowledged.

%===========================
\clearpage
\section*{References}

%\bibliography{/Users/milligen/Documents/Bibtex/Bibtex_database}

\end{document}